\documentclass[%
reprint,
amsmath,amssymb,
aps,
showpacs,preprintnumbers,
superscriptaddress,
prl,
]{revtex4-1}
\usepackage{graphicx}
\usepackage{dcolumn}
\usepackage{bm}

\begin{document}

%\preprint{}

%\title{Study of the $\rm ^9Be=({}^8Be+\it n)$  cluster structure via the Trojan Horse Method}
\title{Experimental study to explore the $\rm ^8Be$ induced nuclear reaction\\ via the Trojan Horse Method}

\newcommand{\ahu}{Anhui University, Hefei 23061, China}
\newcommand{\brcc}{Beijing Radiation Center, Beijing 100875, China}
\newcommand{\ciae}{China Institue of Atomic Energy, P.O. Box 275(18), Beijing 102413, China}
\newcommand{\gik}{Faculty of Engineering Sciences, GIK Institute of Engineering Sciences and Technology, Topi 23640, Khyber Pakhtunkhwa, Pakistan}
\newcommand{\infn}{Laboratori Nazionali del Sud-INFN, Catania, Italy}
\newcommand{\udc}{Dipartimento di Fisica e Astronomia, Univ. di Catania, Italy}

\author{Wen Qun-Gang}\email{qungang@ahu.edu.cn}\affiliation{\ahu}
\author{Li Cheng-Bo}\email{licb2008@gmail.com}\affiliation{\brcc}
\author{Zhou Shu-Hua}\affiliation{\ciae}
\author{Bakhadir Irgaziev}\affiliation{\gik}
\author{Fu Yuan-Yong}\affiliation{\ciae}
\author{Claudio Spitaleri}\affiliation{\infn}\affiliation{\udc}
\author{Marco La Cognata}\affiliation{\infn}
\author{Zhou Jing}\affiliation{\ciae}
\author{Meng Qiu-Ying}\affiliation{\ciae}
\author{Livio Lamia}\affiliation{\infn}\affiliation{\udc}
\author{Marcello Lattuada}\affiliation{\infn}\affiliation{\udc}
%\thanks{}


\date{\today}

\begin{abstract}
To explore a possible indirect method for $\rm ^8Be$ induced astrophysical reactions, the  $\rm ^9Be=({}^8Be+\it n)$ cluster structure was studied via the Trojan Horse Method.
It is the first time to study a super short life nucleus $\rm ^8Be$ via the Trojan Horse Method, and it is the first time to make a valid test for $l=1$ Trojan-horse  nucleus.
The $\rm ^9Be$ nucleus is assumed to have a ($\rm {}^8Be+\it n$) cluster structure and used as the Trojan-horse nucleus.
The $\rm ^8Be$ nucleus acts as a participant, while the neutron is a spectator to the virtual $\rm ^8Be +{\it d}\rightarrow \alpha + {}^6Li$ reaction via a suitable 3-body reaction $\rm ^9Be +{\it d}\rightarrow \alpha + {}^6Li +\it n$.  
The experimental neutron momentum distribution inside $\rm ^9Be$ was reconstructed.
The agreement between experimental and theoretical momentum distribution indicates that there should be a ($\rm {}^8Be+\it n$) cluster structure inside $\rm ^9Be$.
Therefor the experimental study of $\rm ^8Be$ induced reactions, for example the experimental measurement of the $\rm ^8Be +\alpha \rightarrow {}^{12}C$ reaction proceeding through the Hoyle state, is possible.
%Therefor, the experimental study of $\rm ^8Be$ induced reactions, such as the Hoyle-state in Carbon-12 through the reaction of $\rm ^8Be +\alpha \rightarrow\ ^{12}C^*(Hoyle-state)\rightarrow\ ^8Be +\alpha$, is possible.
\end{abstract}

\pacs{21.60.Gx, 24.10.-i, 24.50.+g, 25.70.Hi}
%\keywords{}

\maketitle

\section{Introduction}
%$\rm ^8Be$ is an interesting nucloen when it is a pair of $\alpha$ particles and has a very short lifetime against $\alpha$ decay.
%But its reactions would not be neglected in the environment of nuclear astrophysics. 
Nuclear reactions induced by $\rm ^8Be$ are very important for the study of astrophysical nucleosynthesis, especially for the production of C-12 via the Hoyle state, and provide a new possible way to solve the lithium puzzles in nuclear astrophysics.

The production of the element carbon is one of the key reactions in stellar nucleosynthesis.
Ordinarily, the probability of the triple $\alpha$ process is extremely small.
The most abundant isotope, $\rm ^{12}C$, is created through the formation of the $\rm ^8Be$ ground state as intermediate state.
However, $\rm ^8Be$ has a very short lifetime against $\alpha$ decay, implying that it should almost always break up rather than become the seed for carbon.
The astrophysicist Fred Hoyle recognized that the observed abundance requires an accelerating mechanism and he predicted the existence of a excited state in $\rm ^{12}C$ close to the threshold for $\rm\ ^8Be +\ ^4He$ fusion \cite{Hoyle1953}.
This resonance greatly increases the probability that an incoming $\alpha$ will combine with $\rm ^8Be$ to form $\rm ^{12}C$.
The excited state, with $\rm E^* = 7.6 MeV$, $J^{\pi} = 0^+$, indeed exists, and is often referred to as the "Hoyle state".
But, there is no way to directly study the reaction $\rm ^8Be +\alpha \rightarrow\ ^{12}C^*(Hoyle-state)\rightarrow\ ^8Be +\alpha$ up to now.

To solve the lithium puzzle in nuclear astrophysics, there will be a new possible way when taking into account  the $\rm ^8Be$ induced reactions.
The observed $\rm ^6Li$ abundance in low-metallicity stars is about three orders of magnitude large than that predicted from standard big bang nucleosynthesis.
In addition, $\rm ^7Li$ observations lie a factor 3-4 below the big bang prediction \cite{FieldsLiProblem,0004-637X-644-1-229}.
%In addition, the inferred $\rm ^7Li$ primordial abundance is three times smaller than the big bang prediction \cite{FieldsLiProblem,0004-637X-644-1-229}.
%Assuming there was a period of time in primordial nucleosynthesis, when the temperatures and densities were suitable that $\rm ^8Be$ was produced faster than it decayed.
The fusion of two $\alpha$ into $\rm ^8Be$ requires 0.093 MeV. 
While there is a resonance in $\rm ^{10}B$ at about 0.1 MeV above the $\rm\ ^8Be +\ ^2H$ fusion threshold.
So this resonance maybe markedly increase the $\rm ^6Li$ production by reaction $\rm\ ^8Be +\ ^2H\rightarrow\ ^4He +\ ^6Li$, when the temperatures and densities were suitable in some astrophysical environment.
On the other hand, the $\rm ^8Be +\ ^7Li$ reaction proceeding maybe have effect for destroying $\rm ^7Li$.
%At the same time, the $\rm ^7Li$ production through $\rm ^4He +\ ^3H\rightarrow\ ^7Li + \gamma$  would be suppressed because a large number of $\alpha$ had been combined into $\rm ^8Be$. 

Although the nuclear reactions induced by $\rm ^8Be$ are so important, however due to the short life of $\rm ^8Be$ which is only $10^{-16}$s, it can neither be used as a beam nor as as a target by current technical means. 
Therefore, up to now, the $\rm ^8Be$ induced nuclear reactions can not be studied directly.

The Trojan-horse method (THM) \cite{Baur1986135} provides a possible way for indirect research of $\rm ^8Be$ induced nuclear reactions by using $\rm ^9Be$ as the Trojan horse nucleus, assuming that $\rm ^9Be$ has a ($\rm ^8Be$ + n) cluster structure \cite{Seely2009, PhysRevC.84.031601} with the orbital angular momentum $l=1$.
 
The THM  is an indirect approach to determine the energy dependence of $S$ factors of astrophysically relevant two-body reactions.
The aim of the THM is to extract the cross-section of an astrophysically relevant two-body reaction
\begin{equation}
	A+x\rightarrow C+c
\end{equation}
from a suitably chosen reaction
\begin{equation}
	A+a\rightarrow C+c+b
\end{equation}
with three particles in the final state assuming that the Trojan-Horse $a$ is composed predominantly of clusters $x$ and $b$.
If the $C+c$ system is considered as an excited state of the compound system $B$.
The formulation Eq.(\ref{eq-23}) is the direct relation of the three-body cross-section to the two-body cross-section for partial wave $l$ after selecting the quasi-free events \cite{PhysRevC.69.055806,PhysRevC.71.058801}. 
\begin{equation}
	\frac{d^3\sigma}{dE_{Cc}d\Omega_{Cc}d\Omega_{Bb}}\propto {\rm KF}\cdot|W(\vec{Q}_{Bb})|^2\cdot\Big(\frac{d\sigma}{d\Omega}\Big)^{\rm HOES}_{c.m.}
	\label{eq-23}
\end{equation}
$\rm KF$ is a kinematical factor containing the final state phase space factor and is a function of the masses, momenta and angles of the outgoing particles.
$|W(\vec{Q}_{Bb})|^2$ is the momentum distribution of the spectator $b$ inside the Trojan-horse nucleus $a$.
The term $\Big(\cfrac{d\sigma}{d\Omega}\Big)^{\rm HOES}_{c.m.}$ is the differential two-body cross section induced at energy $E_{c.m.}$ given in post-collision prescription by $E_{c.m.}=E_{Ax}=E_{Cc}-Q_{2body}$.
The variable $E_{Cc}$ is the relative energy between the outgoing particles and $Q_{2body}$ is the Q-value of the vietual two body reaction.
%%%%%%%%%%%%%%%%%%%%%%%%%%%%%%%%%%%%%%%%%%%%%%%%%%%%%%%%%%
%\begin{equation}
%	\frac{d^3\sigma}{dE_{Cc}d\Omega_{Cc}d\Omega_{Bb}}=KF|W(\vec{Q}_{Bb})|^2P_l\frac{d\sigma_l}{d\Omega}(Ax\rightarrow Cc)
%	\label{eq-23}
%\end{equation}
%Where $KF$ is the kinematical factor
%\begin{equation}
%	KF=\frac{\mu_{Aa}\mu_{Bb}\mu_{Cc}}{(2\pi)^5\hbar^6}\frac{k_{Bb}k_{Cc}}{k_{Aa}}\frac{16\pi^2}{k^2_{Ax}Q^2_{Aa}}\frac{v_{Cc}}{v_{Ax}}\frac{k^2_{Ax}}{k^2_{Cc}}
%\end{equation}
%$P_l$ is the penetration function
%\begin{equation}
%	P_l=k^2_{Ax}R^2z^2_l[F^2_l(\eta_{Ax};k_{Ax}R)+G^2_l(\eta_{Ax};k_{Ax}R)]
%\end{equation}
%and W is the momentum distribution of the spectator $b$ inside the Trojan-horse nucleus $a$
%\begin{equation}
%	W(\vec{Q}_{Bb})=-\Big(\varepsilon_{a}+\frac{\hbar^2Q^2_{Bb}}{2\mu_{xb}}\Big)\Phi_a(\vec{Q}_{Bb})
%	\label{eq-W}
%\end{equation}
%%%%%%%%%%%%%%%%%%%%%%%%%%%%%%%%%%%%%%%%%%%%%%%%%%%%%%%%%%
Using the Eq.(\ref{eq-23}), we can reconstruct the momentum distribution of $|W(\vec{Q}_{Bb})|^2$ via THM.

In this paper, the THM was applied to the $\rm ^9Be + {\it d}\rightarrow \alpha +\ ^6Li +{\it n}$ reaction.
The $\rm ^9Be$ nucleus was assumed as $n + \rm\ ^8Be$ cluster structure and used as the Trojan-horse nucleus.
The $\rm ^8Be$ nucleus acts as a participant, while the neutron is a spectator to the virtual $\rm ^8Be + {\it d}\rightarrow \alpha +\ ^6Li$ reaction.
The experimental neutron momentum distribution inside the Trojan-horse nucleus $\rm ^9Be$ would be reconstructed.
This can give a test if there is a cluster structure of $\rm ^9Be = ({}^8Be + n)$  inside $\rm ^9Be$, which is a feasibility study for the further  research work of $\rm ^8Be$ related  nuclear reactions.
%And it can give a valid test, for the first time, to the Trojan-horse nucleus with the orbital angular momentum $l=1$.
And it is also worth to mention that this is the first experiment to use a $l=1$ nucleus as Trojan-horse nucleus.

\section{experiments}
A beam of $\rm ^9Be$ at 22.4 MeV was provided by the HI-13 tandem accelerator at China Institute of Atomic Energy.
A strip of deuterated polyethylene target $\rm CD_2$ of about $\rm 160\mu g/cm^2$ in thickness and about 1 mm in width was oriented with its surface perpendicular to the beam direction.
Using the strip target limited the horizontal width of the beam spot in 1 mm to decrease the angle error.

A position sensitive detector ($\rm PSD_1$) was placed at $15^\circ\pm 5^\circ$ to the beam line direction and about 240 mm from the target.
In the other side of the beam line, a $\rm DPSD$ (Dual Position Sensitive Detector, consisted of $\rm PSD_u$ in the upside and $\rm PSD_d$ downside) was used at $8.7^\circ\pm 5^\circ$ and about 250 mm distance from the target.
The trigger for the event acquisition was generated by events having particle multiplicity larger than 1.
Energy and position signals for the detected particles were processed by standard electronics and sent to the acquisition system for online monitoring and data storage for offline analysis.

\section{data analysis}
For the angular calibration, an equally spaced grid was mounted in front of each detector and the angular position of each grid inside the scattering chamber was determined by using an optical system.
The position and energy calibrations were performed by means of a standard $\alpha$ source, the $\rm\ ^{12}C(\ ^6Li,\ ^6Li)\ ^{12}C$ and $\rm\ ^{197}Au(\ ^6Li,\ ^6Li)\ ^{197}Au$ scattering with $\rm ^6Li$ beam at 15 and 10 MeV, and the $\rm\ ^{197}Au(\ ^9Be,\ ^9Be)\ ^{197}Au$ scattering with $\rm ^9Be$ beam at 8 and 22.4 MeV.

\begin{figure}
\includegraphics[width=0.46\textwidth]{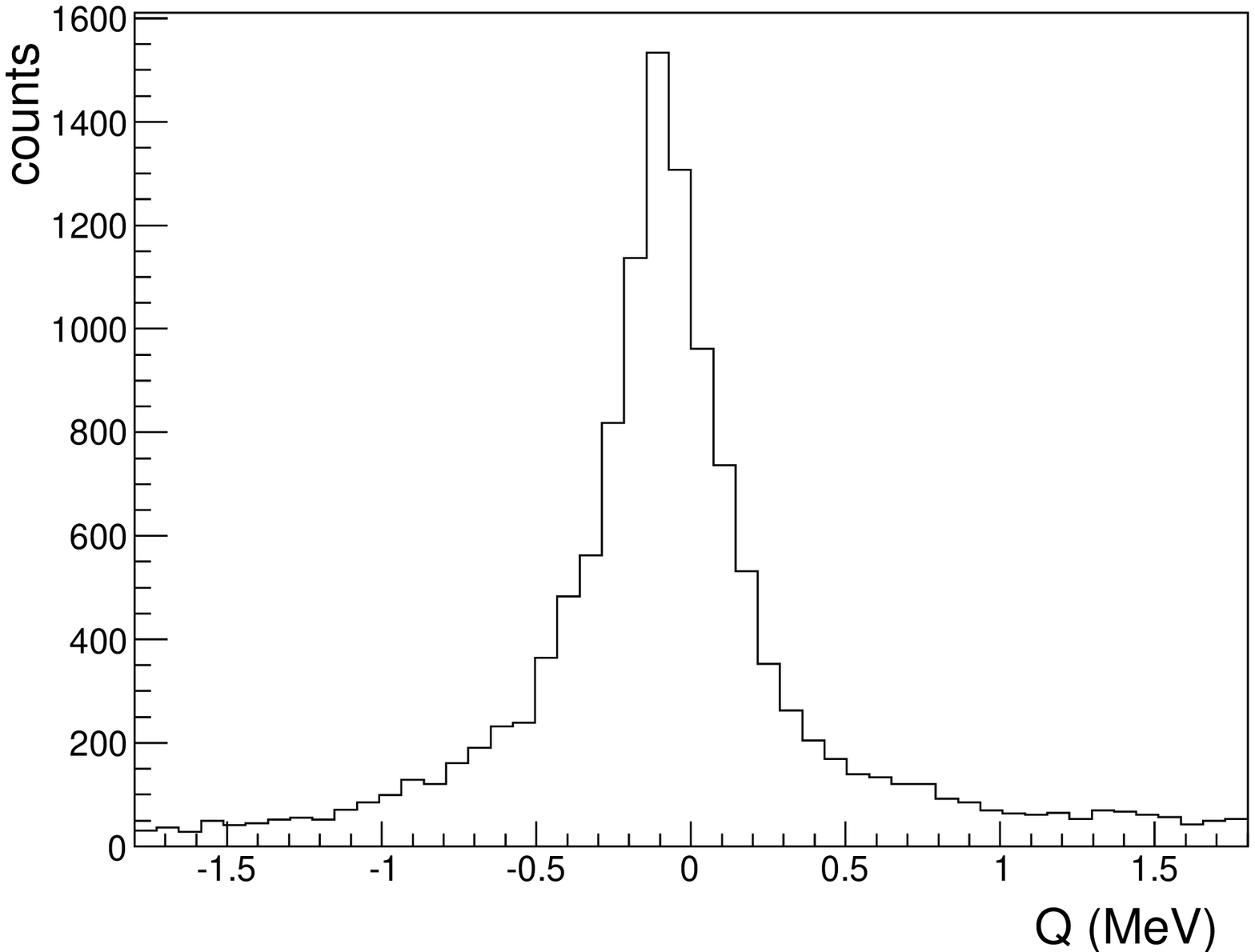}%
\caption{\label{fig-Q} Experimental Q-value for the $\rm ^2H( {}^9Be, \alpha\ ^6Li)\it n$ reaction peaked at $\sim$ -0.1 MeV.}
\end{figure}
After position and energy calibration, the selection of the $\rm\ ^2H(\ ^9Be,\alpha\ ^6Li)\it n$ reaction channel was performed.
We assume all those particles detected by $\rm PSD_1$ are $\rm ^6Li$, all particles detected by $\rm PSD_u$ (in the upside of the $\rm DPSD$) are $\alpha$, and the third particles calculated with $\rm\ ^2H(\ ^9Be, \alpha\ ^6Li)\it n$ three-body reaction kinematic equations are $n$.
In order to understand the experimental data, a set of simulation system was established for the experimental setup based on Geant4 \cite{arXiv1408.6291, JangZJ2014}.
These right events were selected with the help of the simulation system.
The Q-value of the $\rm\ ^2H(\ ^9Be, \alpha\ ^6Li)\it n$ reaction was reconstructed by means of the momentum and energy conservation.
The experimental spectrum of the Q-value is shown in Fig.\ref{fig-Q}, for which a peak centered about -0.1 MeV was obtained in agreement with the expected theoretical value.
The majority of events would be considered in the reaction channel of interest when the Q-value was selected from -0.9 to 0.7 MeV. 

\begin{figure}
\includegraphics[width=0.46\textwidth]{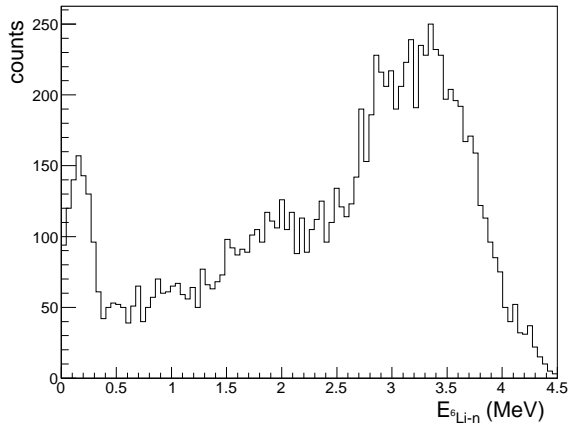}%
\caption{\label{fig-Li7star} Relative energy of $\rm ^6Li-\it n$ spectra. There is a peak about 0.2 MeV belonging to the 7.45 MeV excited states of $\rm ^7Li^*$.}
\end{figure}

\begin{figure}
\includegraphics[width=0.46\textwidth]{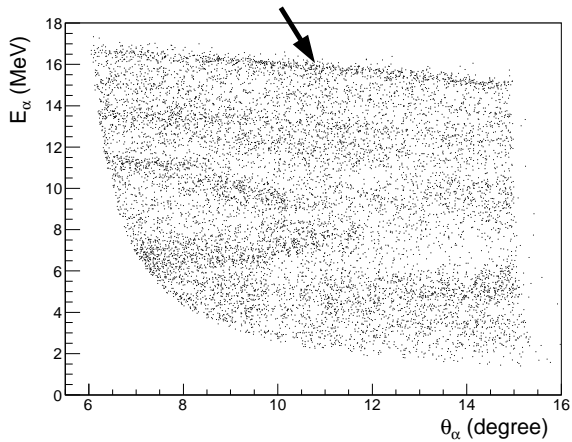}%
\caption{\label{fig-2DLi7star} $\rm E_{\alpha}$ vs $\theta_{\alpha}$ relative two-dimensional plots. The events, which appears a curve and were indicated by an arrow, mainly came from the sequential mechanism reaction $\rm ^9Be +\ ^2H\rightarrow\ ^7Li^* + \alpha\rightarrow \ ^6Li + {\it n} +\alpha$.}
\end{figure}

The next step is the quasifree mechanism identification of the $\rm \alpha+\ ^6Li+\it n$ exit channel.
There is a peak of $\rm E_{\,^6Li-n}=0.2$ MeV in Fig.\ref{fig-Li7star}, which belongs to the 7.45 MeV excited states of $\rm ^7Li^*$.
It means that the events mainly came from the sequential mechanism reaction $\rm ^9Be +\ ^2H\rightarrow\ ^7Li^* + \alpha\rightarrow \ ^6Li + {\it n} +\alpha$. 
Because $\alpha$ came from the two-body reaction $\rm ^9Be +\ ^2H\rightarrow\ ^7Li^* + \alpha$, the events appears a curve in the $\rm E_{\alpha}$ vs $\theta_{\alpha}$ relative two-dimensional plots (see Fig.\ref{fig-2DLi7star}).
So those events, which were around the curve indicated by an arrow in Fig.\ref{fig-2DLi7star}, would be eliminated for the quasifree mechanism identification. 

In this paper, we expect the Trojan-horse nucleus to be $\rm\ ^9Be =\ ^8Be + n$.
But, in fact, the neutron can also come from the Trojan-horse nucleus $\rm\ ^2H = p + n$.
An intermediate process was assumed exist when those events meet the qusifree condition \cite{PRC2008}.
Three criteria for the experimental identification of the qusifree mechanism were introduced \cite{JPG2011}.
Two different Trojan-horse nucleus situation were respectively applied to these criteria.
The events will be selected when they are agree with the Trojan-horse nucleus as $\rm\ ^9Be =\ ^8Be + n$, and be eliminated when they are agree with the Trojan-horse nucleus as $\rm\ ^2H = p + n$.

%The theoretical momentum distribution of the spectator $n$ inside $\rm ^9Be$ is a peak in 0 MeV/c (see Fig.\ref{fig-Be8distribution}). 

\begin{figure}
\includegraphics[width=0.46\textwidth]{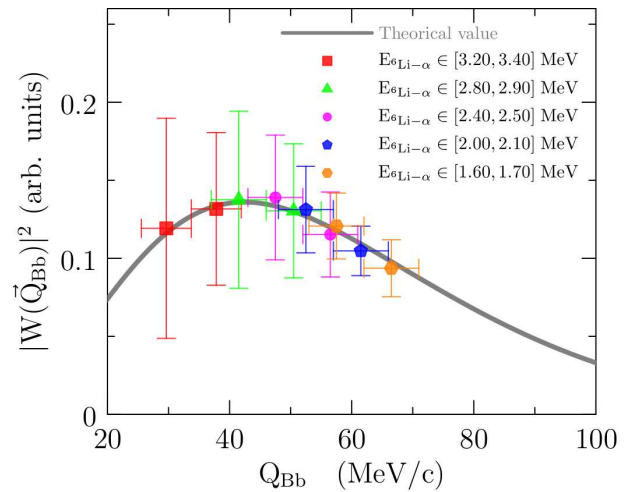}%
\caption{\label{fig-Be8distribution} Comparison between the experimental neutron momentum distribution (different color solid symbols) and theoretical one (grey line). The different color solid symbols come from cutting different range of $\rm ^6Li-\alpha$ relative energy and extracting the neutron momentum distribution. The error bars include only the statistical one.}
\end{figure}
\section{extract of the neutron momentum distribution}
The term $\Big(\cfrac{d\sigma}{d\Omega}\Big)^{\rm HOES}_{c.m.}$ in Eq.\ref{eq-23} represents the nuclear part of the differential cross section for the virtual two-body reaction $\rm ^8Be(\ ^2H,\alpha)\ ^6Li$ that in post-collision prescription occurs at an energy \[E_{c.m.} = E_{\rm ^6Li-\alpha}-Q_{2b},\] where $E_{\rm ^6Li-\alpha}$ is the $\rm ^6Li-\alpha$ relative energy and $Q_{2b}$ is the two-body $Q$ value.

To reconstruct the neutron momentum distribution, a small $\rm ^6Li-\alpha$ relative energy window (about 100 keV) was selected.
In such a small energy window, $\Big(\cfrac{d\sigma}{d\Omega}\Big)^{\rm HOES}_{c.m.}$ can be considered constant.
Thus the experimental $|W(\vec{Q}_{Bb})|^2$ distribution was extracted by dividing the three-body coincidence yield by the kinematic factor.
The results are showed in Fig.\ref{fig-Be8distribution} with the different color solid symbols.

\section{results and discussion}
In Fig.\ref{fig-Be8distribution}, the theoretical neutron momentum distribution was calculated based on assuming $\rm ^9Be$ nucleus as $n+\rm\ ^8Be$ cluster structure with the orbital angular momentum $l=1$. 
To solve the Schr$\rm\ddot{o}$dinger equation we applied the Wood-Saxon potential with the standard values for the radius parameter and diffuseness while the potential well depth was fixed using the $\rm ^8Be-n$ binding energy. 
The orbital angular momentum of neutron was assumed to be unity because $\rm J^{\pi}=\frac{3}{2}^-$ for the ground state of $\rm ^9Be$.
The agreement between experimental and theoretical momentum distribution indicates the $n+\rm\ ^8Be$ cluster structure of $\rm ^9Be$, and represents a very strong check for the existence of the qusifree mechanism in the present data.
%In this work, the corresponding qusifree $\rm ^6Li-\alpha$ relative energy is 3.86 MeV.
%Fig.\ref{fig-Be8distribution} shows that the neutron momentum is small nearby 0 when the $E_{\rm ^6Li-\alpha}$ is near the qusifree energy, and it becomes larger with the $E_{\rm ^6Li-\alpha}$ goes away from the qusifree energy point.

In advance, there will be a possible method to study  $\rm ^8Be$ induced reaction in experiment via THM by using the $\rm ^9Be=(n+\rm\ ^8Be)$ as TH nucleus. 
%The lifetimes of $\rm ^8Be$ is only $10^{-16}$s, so it can not be used as a beam or a target by the current technical means.
One interesting work, studying the nature of Hoyle-state in Carbon-12 \cite{Hoyle1953,Physics.4.94} through the reaction of $\rm ^8Be + \alpha\rightarrow\ ^{12}C^*(Hoyle-state)\rightarrow\ ^8Be +\alpha$ in experiment, will be possible to be done at the next step. 

\begin{acknowledgments}
The authors thank Lin Chengjian's group and Li Xia from CIAE for their kind help during the experiment measurement.
This work is supported by the National Natural Science Foundation of China (11075218, 11321064), the 973 program of China (2013CB834406) and Beijing Natural Science Foundation (1122017). 
This work has been partially supported by the Italian Ministry of University MIUR under the grant RFBR082838(FIRB2008).
\end{acknowledgments}

\nocite{*}
%\bibliography{Be8inBe9}
%

\end{document}